\documentclass[prl,twocolumn,showpacs]{revtex4}
\usepackage{array,amsmath,amssymb,graphicx,pslatex}

\begin{document}

\title{Non-Ohmic variable-range hopping transport in one-dimensional
conductors}

\author{M. M. Fogler and R. S. Kelley}


\affiliation{Department of Physics, University of
California San Diego, La Jolla, California 92093}

\date{\today}

\begin{abstract}

We investigate theoretically the effect of a finite electric field on
the resistivity of a disordered one-dimensional system in the
variable-range hopping regime. We find that at low fields the transport
is inhibited by rare fluctuations in the random distribution of
localized states that create high-resistance ``breaks'' in the hopping
network. As the field increases, the breaks become less resistive. In
strong fields the breaks are overrun and the electron distribution
function is driven far from equilibrum. The logarithm of the resistance
initially shows a simple exponential drop with the field, followed by a
logarithmic dependence, and finally, by an inverse square-root law.

\end{abstract}
\pacs{72.20.Ee, 72.20.Ht, 73.63.Nm}

\maketitle

\noindent{\it Motivation and main results\/}.--- Potential applications
in nanotechnology has generated a widespread interest in one-dimensional
(1D) conductors, including carbon anotubes, nanowires, and conducting
molecules. Before such applications may become practical, a number of
fundamental physics questions need to be resolved. One of them is the
effect of disorder on transport properties. Despite a great progress in
nanofabrication, the disorder is hard to get rid off, especially if a
low-cost processing is desired. The disorder causes localization of
electron states, so that the low-temperature transport in 1D systems is
often of the hopping type~\cite{Mott_book, ES84}. That is why a new look
at the theory of a 1D hopping conduction may be timely. One of the open
questions here is the mechanism of the non-Ohmic hopping transport,
observed in a number of recent experiments~\cite{Tang_02, Aleshin_04,
Tzolov_04}. Especially interesting and challenging is the ultimate
low-$T$ limit where the \emph{variable-range} hopping (VRH), i.e.,
hopping between distant localized states dominates. Remarkably, this
problem has long remained untouched by theorists except for one early
numerical study~\cite{McInnes_90}. Recently, however, the following
analytical formula has been proposed~\cite{Nattermann_03}: 
\begin{equation}
\rho \propto \exp
\left(C {k_B T_0} / {F a}\right)^{1/2},
\label{rho_strong}
\end{equation}
where $\rho$ is the resistivity, $F = e {\cal E}$ is the product of the
electron's charge and the electric field, $a$ is the localization
length, $T_0$ a parameter of dimension of temperature defined below, and
$C$ is a numerical coefficient left undetermined. This formula has
immediately become a matter of controversy because in its derivation a
very important and nontrivial peculiarity of the 1D geometry was
overlooked: unavoidable highly resistive segments (``breaks'') on the
conducting path~\cite{Kurkijarvi_73, Lee_84, Raikh_89}. In an attempt to
include this physics, Eq.~(\ref{rho_strong}) was subsequenty revised in
a follow-up article~\cite{Malinin_04}. Yet the actual situation is more
intriguing. As demonstrated below, the problem has quite a rich regime
diagram. The resistivity may take three different functional forms,
depending on the strength of the electric field (weak, intermediate, or
strong). It is Eq.~(\ref{rho_strong}) that proves to be correct in
strong fields, $F \gg {k_B T}/{a}$, not its subsequent
revision~\cite{Malinin_04}. We compute the exact coefficient in
Eq.~(\ref{rho_strong}) to be $C = 8$. The main message of our analysis
is that the role of the breaks diminishes and eventually becomes
insignificant as the electric field increases. We hope that these ideas
and new results may provide a much needed theoretical basis for
the purposeful experimental study of the non-Ohmic VRH expected
to be reported in the near future.

\noindent{\it Model\/}.--- Since the problem has proved to be rather
delicate, in this study we approach it within a simplified model of {\it
non\/}-interacting electrons. (In real systems interaction effects can
be suppressed by placing the 1D conductor near a metallic gate.) The
localized states are regarded as random points $(x_j, \varepsilon_j)$
with the Poisson distribution of a constant density $g$ in the
position-energy space. This is justified at low $T$ where electron
transport is determined by a narrow range of energies. Similarly, at low
enough $T$ we can neglect the field-dependence of the localization
length $a$ because the electric fields of interest, say, $F \sim k_B T /
a$, are small.

The rate of hopping $\Gamma_{j k}$ from site $j$ to site $k$ is given
by the usual model form~\cite{Mott_book, Shklovskii_76}
\begin{equation}
\frac{\Gamma_{k j}}{\gamma}  = f_j (1 - f_k) \exp\!\left[
-\frac{2}{a}|x_j - x_k| - \frac{\max(D_{j k}, 0)}{k_B T}\right],
\label{I_kj}
\end{equation}
where $\gamma$ of dimension of frequency is determined by the
electron-phonon coupling, $0 < f_j < 1$ is the occupation factor of the
$j$-th site, and $D_{k j} \equiv \varepsilon_k - \varepsilon_j - F (x_k
- x_j)$. The net current from $j$ to $k$ is given by $I_{k j} =
e(\Gamma_{k j} - \Gamma_{j k})$. The transport problem amounts to
solving the current conservation equations $\sum_{k \neq j} I_{k j} = 0$
for $f_j$'s at a given fixed $F$. One conventional way to present the
result is in the form $f_j = \{\exp[(\varepsilon_j - \mu_j) / k_B T] +
1\}^{-1}$, where $\mu_j$ is termed the chemical potential of the $j$-th
site. In the most relevant situation, $\Delta_{k
j} \equiv |\varepsilon_j - \mu_j| + |\varepsilon_k - \mu_k| +
|\varepsilon_j - \varepsilon_k| \gg k_B T$, the currents $I_{k j}$ are
given by~\cite{Pollak_76} 
\begin{equation}
I_{k j} \simeq e\gamma \exp\!\left(
-\frac{2}{a}|x_k - x_j| - \frac{\Delta_{k j}}{2 k_B T}\right)
\sinh\frac{\xi_j - \xi_k}{k_B T},
\label{I_pair}
\end{equation}
where $\xi_j = \mu_j - F x_j$ is the electrochemical potential. Knowing
$\mu_j$'s (or $f_j$'s), one then computes the total current $I$ and the
resistivity $\rho(F, T) = e I / F$.

\begin{figure}
\centerline{
\includegraphics[width=3.0in,bb=213 598 451 716]{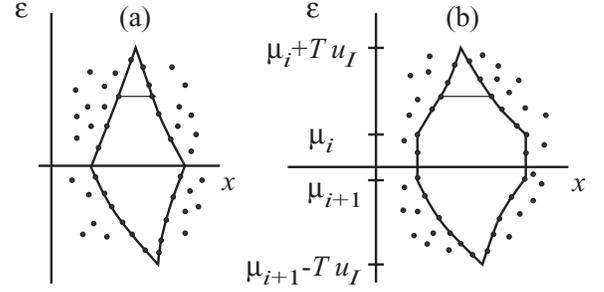}
}
\vspace{0.1in}
\caption{
(a) Ohmic and (b) non-Ohmic breaks. The dots represent hopping sites.
The horizontal thin lines are the links of the optimal path carrying
the current across the breaks. The width of each break is $a u_I / 2$.
\label{Fig_breaks}
}
\end{figure}

\noindent{\it Ohmic regime\/}.--- In general, the above transport
equations are nonlinear in $\mu_j$ but in the Ohmic limit, $F \to 0$, $T
> 0$, both $\mu_j$ and $\xi_j - \xi_k$ are smaller than $T$ and
Eq.~(\ref{I_pair}) becomes $(\xi_j - \xi_k) / e = I_{k j} R_{k j}$,
where $R_{j k} = R_0 \exp [(2 / a)|x_k - x_j| + (\Delta_{k j} / 2 k_B
T)]$. Therefore, the system is equivalent to the network of resistors
$R_{j k}$, while $\xi_j / e$ is simply the voltage at the
$j$-node~\cite{Mott_book,ES84}. Here $R_0 = k_B T / (e^2 \gamma)$ and
$\Delta_{j k}$ is to be computed at $\mu_j = \mu_k = 0$. The resistances
$R_{k j}$ differ by many orders of magnitude from one another and this
is what makes the problem nontrivial. Let us review the basic ideas
developed for its solution. A crucial step~\cite{Mott_book,ES84}
is to realize that among many possible current paths through
the network, a single optimal one heavily dominates the conduction.
In 1D this path has the topology of a linear chain, so that we can label
the nodes of the path consequtively: $1, 2, \ldots, i, i + 1$, {\it
etc\/}. The typical separation in position ($x_M$) and in energy
($\varepsilon_M$) between these nodes follows from the
conventional Mott's argument~\cite{Mott_book, ES84}:
\begin{equation}
\frac{\varepsilon_M(T)}{k_B T} =
\frac{x_M(T)}{a} = u_M \equiv \sqrt{\frac{T_0}{T}},\:\:
k_B T_0 \equiv \frac{1}{g a}.
\label{x_M}
\end{equation}
Thus, the logarithm of the resistance of a typical link is $\ln R_{i, i
+ 1} \sim x_M / a + \varepsilon_M / k_B T \sim u_M$. In addition to
typical links, there are places in the conductor where hopping sites are
locally absent in a large area $A$ of the $(x, \varepsilon)$-plane. The
probability of every such ``break'' is very small, $\sim \exp(-g A)$,
but its resistance far exceeds that of a typical link. A careful
analysis of Raikh and Ruzin (RR)~\cite{Raikh_89} showed that the
dominant contribution to $\rho$ is provided by breaks that are shaped as
diamonds and have the sizes $u_\Omega a / 2$ and $2 T_0$ in the position
and energy dimensions, respectively, see
Fig.~\ref{Fig_breaks}(a)~\cite{Comment_on_diamond}. Here $u_\Omega = T_0
/ T$. These ``optimal'' breaks have resistances $R_\Omega = R_0
\exp(u_\Omega)$ and are present in concentration $N_\Omega \sim
\exp(-u_\Omega / 2) / x_M$, which yields $\rho \sim R_\Omega N_\Omega$,
i.e.,~\cite{Raikh_89}
\begin{equation}
        \ln \rho(0, T) \simeq T_0 / 2 T,\quad T \ll T_0.
\label{rho_activated}
\end{equation}
As customary in this field, we will not attempt to compute the
pre-exponential factor of the resistivity~\cite{Comment_on_prefactor}.

If the 1D Ohmic VRH seems like a nontrivial problem, it becomes even
more so in the {\it non\/}-Ohmic regime we wish to discuss next. In
particular, we address the following new set of questions: What is the
highest $F$ when Eq.~(\ref{rho_activated}) still applies? Do breaks
continue to play a role at larger $F$? How does $\rho$ depend on $F$ at
such fields? Our answers to these questions are given below.

%
%

\noindent{\it Intermediate fields\/}.--- We begin by showing that the
region of validity of Eq.~(\ref{rho_activated}) is very limited. Indeed,
in the derivation it was assumed that the voltage drop across each link
is less than $k_B T / e$. However the voltage drop on every optimal break
is equal to $I R_\Omega = (F / e \rho) R_\Omega = F / e N_\Omega$; therefore,
Eq.~(\ref{rho_activated}) is justified only at $F \ll F_\Omega \sim (k_B
T / x_M) \exp(-T_0 / 2 T)$. To compute $\rho$ at larger $F$ a different
approach is needed. Our strategy resembles that of
Ref.~\cite{Shklovskii_76}. We assume that a given fixed current $I$
flows through the system and compute $F = \langle d \xi / d x \rangle$.
Then we find $\rho$ from the relation
\begin{equation}
\ln\left[\frac{\rho(F, T)}{(R_0 / a)}\right] = u_I -
\ln\left(\frac{k_B T}{F a}\right), \quad u_I \equiv
\ln \left(\frac{e\gamma}{I}\right).
\label{rho_from_u_I}
\end{equation}
At this stage we focus on the intermediate field (current) regime,
defined by $u_M \ll u_I \ll u_\Omega$. Let $\Delta\xi_i \equiv \xi_i -
\xi_{i + 1}$ be the electrochemical potential difference across the
$i$-th link of the optimal path. Since $x_M$ is the correct estimate of
the {\it typical\/} length of the link, we have $F \sim \langle \Delta
\xi_i \rangle / x_M$, where the angle brackets denote averaging over
$i$. Introducing $P_\xi$, the probability distribution function (PDF) of
the random quantity $\Delta \xi_i$, we can write this as
\begin{equation}
F \sim \frac{1}{x_M} \int\limits_0^\infty P_\xi(\Delta \xi)
\Delta \xi d \Delta\xi.
\label{F_from_P}
\end{equation}
Let us establish the functional form of $P_\xi$. The behavior of
$P_\xi(\Delta\xi)$ at $\Delta\xi \ll k_B T$ is determined by links with
Ohmic resistances $R_0 e^u$, where $u \ll u_I$. They remain operating in
the Ohmic regime, and so for them $\Delta\xi_i / k_B T = \exp(u - u_I) \ll
1$. Using RR's results for the PDF of the (diamond-shaped) Ohmic breaks,
we obtain
\begin{equation}
P_\xi = \frac{C}{\Delta\xi} \exp\left\{-\frac{T}{2 T_0}
\left[u_I - \ln \left(\frac{k_B T}{\Delta\xi}\right)\right]^2\right\},
\label{P_xi_Ohmic}
\end{equation}
where $1 \ll \ln (k_B T / \Delta\xi) < u_I - u_M$ and $C$ is a prefactor
with a subleading dependence (presumably, a logarithmic one) on
$\Delta\xi$ and $T$. Note that the typical links, $u \sim u_M$, are
included in Eq.~(\ref{P_xi_Ohmic}) [they correspond to the lowest
$\Delta\xi \sim k_B T \exp(u_M - u_I)$ at which this formula is still
valid]. Consider now the links with $\Delta \xi_i \gg k_B T$, which are
necessarily atypical, i.e., some sort of breaks. The geometry of such
non-Ohmic breaks can be established as follows. In a vicinity of a
non-Ohmic break the largest voltage variation is across this break;
therefore, the electrochemical potential has nearly the same value, say,
$\xi_i$, on the sites adjacent to the left boundary of the break and
another value, $\xi_{i + 1}$ for those on the right side. Assuming that
$\Delta\xi_i \simeq \Delta\mu_i$, (borne out by the results) we see from
Eq.~(\ref{I_pair}) that the current through a pair of sites $j$ and $k$
located on the opposite sides of the break is $I_{k j} \simeq e\gamma
\exp [-(2 / a) |x_k - x_j| - (\Delta_{k j} / 2 k_B T) + (\Delta\mu_i /
k_B T)]$. The shape of the non-Ohmic break should be such that the
inequality $I_{k j} \leq I$ is obeyed. The optimal, i.e., the most
probable breaks are those of minimal area. Using only elementary
geometrical considerations one readily finds that the optimal shapes
comprise a continuum of curvilinear hexagons of area $A_i = a u_I (k_B T
u_I + \Delta\xi_i) / 2$, see
Fig.~\ref{Fig_breaks}(b). Hence,
$P_\xi(\Delta\xi_i) \sim \exp(-g A_i)$ for $\Delta\xi_i \gg k_B T$, or,
more accurately, 
\begin{equation}
P_\xi = \frac{C}{k_B T}
\exp\left(-\frac{T u_I^2}{2 T_0} - \frac{u_I \Delta\xi}{2 k_B T_0}\right),
\quad \Delta\xi \gg k_B T.
\label{P_xi_non-Ohmic}
\end{equation}
Equations~(\ref{P_xi_Ohmic}) and (\ref{P_xi_non-Ohmic}) match by
the order of magnitude at $\Delta\xi \sim k_B T$. Substituting them
into Eq.~(\ref{F_from_P}), we find the the main voltage
generators in the system are the {\it non\/}-Ohmic breaks with
$\Delta\xi \sim k_B T_0 / u_I$. For $F$ we obtain
\begin{equation}
F \sim (C k_B T_0 / a u_I^2) \exp (-u_I^2 / 2 u_M^2).
\label{F_non-Ohmic}
\end{equation}
Solving this for $u_I$ and using Eq.~(\ref{rho_from_u_I}), we arrive
at
\begin{equation}
\ln\rho \simeq \sqrt{\frac{2 T_0}{T} \ln \left(\frac{\nu k_B T}{F a}\right)}
          -  \ln \frac{k_B T}{F a},
\quad F_\Omega \ll F \ll \frac{k_B T}{a},
\label{rho_intermediate}
\end{equation}
where $\nu = C u_I^2 / u_M^2$. The main message of this analysis is that
atypically large resistors are progressively eliminated as $F$ increases
($u_I$ decreases). Note that at $u_I \sim u_M$ the
distinction between breaks and typical links disappears, so that the
result $\Delta\xi_i \sim k_B T_0 / u_M$ for the breaks is a good
estimate of the overall $\langle\Delta\xi \rangle$. Thus, at this point $F
\sim k_B T / a$ and therefore $\nu \sim C \sim 1$.

\noindent{\it Strong fields\/}.--- A quick if unilluminating derivation
of $\rho$ at $F \gg k_B T / a$ assumes that Eq.~(\ref{F_non-Ohmic})
holds throughout the non-Ohmic regime. In strong fields, where $u_I \ll
u_M$, the exponential in Eq.~(\ref{F_non-Ohmic}) drops out leaving one
with $F \sim C k_B T_0 / a u_I^2$. Combined with
Eq.~(\ref{rho_from_u_I}), this entails Eq.~(\ref{rho_strong}). Notably,
this formula differs from Eq.~(4.29) of Ref.~\cite{Malinin_04}, by a
large logarithmic factor in the exponential. In view of this discrepancy
(which has a conceptual importantce, cf.~the introduction), we present a
rigorous derivation of Eq.~(\ref{rho_strong}) that clarifies the physics
involved and yields the exact result $C = 8$ for the coefficient $C$.

In strong fields only the forward hops need to be
considered~\cite{Shklovskii_72, Pollak_76}, so that for the $i$-th link
of the optimal path we have $I = e \Gamma_{i + 1, i}$ or, equivalently
[cf.~Eq.~(\ref{I_kj})]
\begin{align}
f_i (1 - f_{i + 1})& \exp[u_I - (2 / a) |x_{i + 1} - x_i|] = 1,
\label{f_i_equation}\\
\epsilon_i \geq \epsilon_{i + 1},&\quad \epsilon_i \equiv
\varepsilon_i - F x_i.
\label{epsilon_i}
\end{align}
A necessary condition for the existence of a physically acceptable
solution, $0 < f_i < 1$, of Eq.~(\ref{f_i_equation}) is
\begin{equation}
               |x_{i + 1} - x_i| < x_I \equiv a u_I / 2
\label{length_constraint}
\end{equation}
(thus, the typical hopping length is field-dependent). A sufficient
condition is known from the theory of a Disordered Asymmetric Simple
Exclusion Process~\cite{Harris_04}, which is described by the identical
equations. It is just a bit more stringent, $|x_{i + 1} - x_i| \leq x_I
- a \ln 2$. Since we are interested in the case $u_I \gg 1$, it is a
legitimate approximation to assume that the
condition~(\ref{length_constraint}) is both necessary and sufficient.
Only the {\it existence\/} of the solution matters, not the actual
values of $f_i$. (Those will be commented on at the end.) We expect on
physical grounds that the optimal path does not deviate indefinitely
from the equilibrium chemical potential, $\sup |\varepsilon_i| <
\infty$. Then the standard percolation theory argument~\cite{Pollak_76}
entails
\begin{equation}
F = \min \lim_{i \to \infty} [(\epsilon_1 - \epsilon_i) / (x_i - x_1)].
\label{F_equation}
\end{equation}
Indeed, if $F$ were smaller than the right-hand side of this equation,
then the contiguous path that obeys the constraints~(\ref{epsilon_i})
and (\ref{length_constraint}) would not exist. If $F$ were larger, the
chosen path would be short-circuited by another one. To compute $F$ from
Eq.~(\ref{F_equation}) we use the fact that statistics of the random
points $(x_j, \epsilon_j)$ is again an uncorrelated Poisson distribution
of density $g$.

%
%
\begin{figure}
\centerline{
\includegraphics[width=2.6in,bb=215 615 391 703]{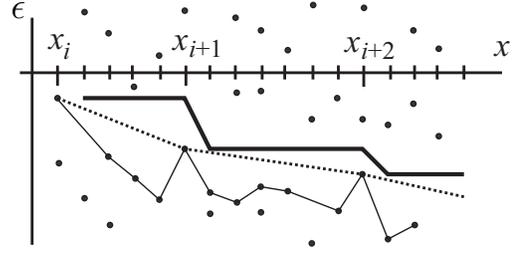}
}
\vspace{0.1in}
\caption{
Plots of $E(x)$ (thin line), $E_<(x)$ (thick line) and the optimal path
(dotted line). Ticks and dots represent lattice sites and
their energy levels, respectively; $x_I = 9 b$ is assumed.
\label{Fig_E}
}
\end{figure}

Consider the following lattice version of the problem. Let all
coordinates be restricted to integer multiples of $b \ll x_I$. On every
site $x_j = j b$ let there be a discrete set of energy levels
$\epsilon_n(x_j)$, $n \in \mathbb{Z}$, drawn from an independent Poisson
distribution with the average energy spacing $1 / g b$. Suppose next
that $\{x_1, \epsilon_0(x_1)\}$ is the starting (the leftmost) site of
the sought optimal path. Denote by $E(x)$ the largest $\epsilon_n(x)$
reachable from this site by a path that obeys the
constraints~(\ref{epsilon_i}) and (\ref{length_constraint}), then $F =
\lim_{x \to \infty} \{[\epsilon_0(x_1) - E(x)] / (x - x_1)\} = -\lim_{x
\to \infty} [E(x) / x]$. Function $E(x)$ satisfies the obvious
recurrence relation: $E(x)$ is equal to the largest $\epsilon_n(x)$ that
still does not exceed $E_<(x)$, where $E_<(x) \equiv \max E(y)$, $x - x_I
< y < x$. This relation is illustrated graphically in
Fig.~\ref{Fig_E}. It is easy to see that function $E_<(x)$ is step-like,
i.e., that there is an increasing sequence of coordinates $x_i$, such
that $E_<(x) = E(x_i) = {\rm const}$ for $x_i < x \leq x_{i + 1}$.
Figure~\ref{Fig_E} aids in observation that given $x_i$, the next site
in this sequence, $x_{i + 1}$, is simply the one with the largest $E(y)$
less than $E_<(x_i)$ among all the points that belong to the vertical
strip $x_i < y < x_i + x_I$. These $x_i$ are in fact the coordinates of
the optimal path (more precisely, they become such once one further
optimizes with respect to the path's starting point). By virtue of this
property, $\Delta E_i = E_<(x_{i}) - E_<(x_{i + 1})$ and $\Delta x_i
= x_{i + 1} - x_i$ are uncorrelated random numbers with average values
$2 k_B T_0 / u_I$ and $a u_I / 4$, respectively (in the limit $b \ll
x_I$). This entails $F = \langle \Delta E_i \rangle / \langle \Delta x_i
\rangle = C k_B T_0 / a u_I^2$ with $C = 8$.

It is worthwhile to clarify why Eq.~(\ref{rho_strong}) is not
invalidated by inevitable large voids in the distribution of sites in
the $(x, \epsilon)$-plane. Whenever the optimal path meets such a void,
$\Delta E_i$ becomes much larger than its average value. However, the
corresponding contribution to $F$ grows linearly with the size of the
void whereas the probability of the void drops exponentially.
These rare large voids have thus a negligible effect on $F(u_I)$.

The derivation above indicates that the site occupations in the energy
band $|\varepsilon_j| \alt k_B T_0 / u_I \sim (k_B T_0 F a)^{1 / 2}$
that carries the current, deviate so much from their equilibrium values,
$\ln (1 / f_j - 1) = \varepsilon_j / k_B T$, that $T$ plays no vital
role in the transport. This agrees with the early result of
Shklovskii~\cite{Shklovskii_72} that at large $|\varepsilon_j|$ the real
temperature $T$ in the last equation gets replaced by the ``effective''
one, $T_{\rm eff} = F a / 2 k_B$. However, at $\varepsilon_j$'s of
relevance to VRH there exist large corrections, $k_B T_{\rm eff} \ln (1
/ f_j - 1) - \varepsilon_j \sim (k_B T_0 F a)^{1 / 2}$, to the
asymptotic large-$\varepsilon$ behavior that prohibit one to derive
Eq.~(\ref{rho_strong}) via a simple replacement $T \to T_{\rm eff}$ in
the weak-field formula, Eq.~(\ref{rho_activated}).

\noindent{\it Experimental implications\/}.--- The length $L$ of real 1D
conductors is seldom very large. Because of that their resistivity in a
VRH regime fluctuates strongly from one sample to the next, depending on
what configuration of breaks is realized~\cite{Lee_84, Raikh_89}. Here
we discuss only the ensemble average value of $\ln\rho$. For the Ohmic
case, in the most common situation $1 \ll {\cal L} \ll T_0 / T$, where
${\cal L} \equiv \ln(L / x_M)$, the result for this quantity is known to
be~\cite{Lee_84, Raikh_89} $\langle\ln \rho(0, T) \rangle \simeq \sqrt{2
{\cal L} T_0 / T}$. In this regime practically all the applied voltage
$V = F L / e$ drops on the single largest break~\cite{Raikh_89}. Hence,
the Ohmic behavior lasts as long as $e V < k_B T$. In some range of $e
V$ above $k_B T$, the same break remains the bottleneck for the
transport. Using Eq.~(\ref{I_pair}) and averaging over the position of
the break along the chain, we obtain the exponential dependence,
\begin{equation}
\ln \langle{\rho}/{\rho(0, T)}\rangle \simeq
 -{e V}/{2 k_B T},\quad k_B T \ll e V \ll e V_L.
\label{exponential_L}
\end{equation}
The upper limit $V_L$ where this formula is valid can be estimated from
the condition $e V_L \sim \langle \Delta \xi \rangle \sim k_B T_0 /
u_I$, which gives $e V_L \sim k_B (T T_0 / {\cal L})^{1/2}$. At $V \gg
V_L$ the infinite-chain results, Eqs.~(\ref{rho_intermediate}) and
(\ref{rho_strong}), apply.

On a qualitative level, existing experimental data are consistent with
our theory. For example, a characteristic S-shaped $\log I-\log V$
curve~\cite{Aleshin_04} follows naturally from our equations. To
unabiguously verify them an earnest effort and a thoughtful experimental
design are required. At present, our knowledge and degree of control
over disorder and other parameters of 1D systems remain poor; therefore,
a wide dynamical range of data is needed to compensate for many
uncertainties. The most suitable for this purpose are systems where
localization length is not too short, e.g., high-quality carbon
nanotubes. Their resistance has to be measured over a broad range of low
$T$, where $\rho$ is determined by the exponential terms we computed
rather than some power-law prefactors we left undetermined here. A care
should be taken to remain in the validity domain of the theory, $L \gg
x_M \sim a \sqrt{T_0 / T}$. The parasitic effects of contact
resistance~\cite{Tang_02, Tzolov_04} need to be done away with reliably.
If the system is quasi-1D rathen than 1D, hopping across the
chains~\cite{Aleshin_04, Tzolov_04} must be negligible. Finally, a
careful statistical analysis of the data is likely to be
needed~\cite{Hughes_96}.

An important qualitative prediction of our theory is that enormous
sample-to-sample variability of the low-$T$ hopping resistance becomes
less of an issue in the non-Ohmic regime. This may facilitate both the
experimental studies we call for and the future theoretical work that
would be needed to include the effects of Coulomb interaction into the
model. Such coupled experimental and theoretical studies will likely to
shed much light on the intriguing transport properties of
low-dimensional conductors.

\noindent{\it Acknowledgements\/}.--- This work is supported by the C.
\& W.\ Hellman Fund, the A. P. Sloan Foundation, and UCSD ASCR. We thank
B.~Fetscher for participation in the early stage of this project and
B.~Shklovskii, T.~Nattermann, and especially M.~Raikh for discussions.



\begin{thebibliography}{99}
\bibitem{Comment_on_prefactor} The calculation of the prefactor is a
difficult open problem~\cite{ES84} while the result is likely to be
nonuniversal.
\bibitem{Comment_on_diamond} In Ref.~\cite{Raikh_89} only symmetric
diamonds were considered.
\end{thebibliography}
\end{document}